\documentclass[
  aps,
  prx,
  reprint,
  amsmath,amssymb,
  floatfix,
  superscriptaddress,
  nofootinbib
]{revtex4-2}

\usepackage{mathtools}
\usepackage{bm}
\usepackage{bbm}
\usepackage{hyperref}

\newcommand{\Id}{\mathbbm{1}}
\newcommand{\ket}[1]{\lvert #1 \rangle}
\newcommand{\bra}[1]{\langle #1 \rvert}
\newcommand{\ii}{\mathrm{i}}
\newcommand{\ee}{\mathrm{e}}
\newcommand{\dd}{\mathrm{d}}
\newcommand{\Tr}{\operatorname{Tr}}
\newcommand{\comm}[2]{\left[ #1 , #2 \right]}
\newcommand{\Hilb}{\mathcal{H}}
\newcommand{\Lop}{L}
\newcommand{\Mop}{M}
\newcommand{\Aop}{A}
\DeclareMathOperator{\spec}{spec}

\begin{document}

\title{Quantum Dynamics from Lax Pair Theory: \\ A Reconstruction from Spectrum Preservation}

\author{P\'eter Szab\'o}
\email{peter88szabo@gmail.com}
\affiliation{Belgian Institute for Space Aeronomy (BIRA-IASB), Brussels, Belgium}
\affiliation{Department of Chemistry, KU Leuven, Leuven, Belgium}

\date{\today}

\begin{abstract}
We reconstruct unitary quantum dynamics from a minimal axiomatic foundation built on Hilbert-space observables and isospectral evolution. The only dynamical assumption is that physical time evolution is a continuous one-parameter flow of Hermitian observables that preserves their spectra, i.e. the possible outcomes of measurement. We show that this assumption is already sufficient to force the Lax form of quantum dynamics. The Heisenberg equation, the time-dependent and time-independent Schrödinger equations, conservation laws, and good quantum numbers then follow as theorems rather than postulates. In this formulation, Lax pair theory supplies the missing dynamical bridge between the measurement structure of a Hilbert space and standard quantum evolution: the Hamiltonian is not assumed, but emerges as the generator required for an isospectral observable flow.
\end{abstract}

\maketitle

\section{Introduction}

Quantum mechanics is an axiomatic theory. In the formulations that descend
from Dirac~\cite{dirac} and von~Neumann~\cite{vonneumann}, its entire content
is carried by a handful of postulates: a physical state is a ray in a complex
Hilbert space; observables are self-adjoint operators whose spectra are the
admissible outcomes of measurement; the Born rule assigns probabilities to
those outcomes; the state evolves in time by the Schr\"odinger equation, a
one-parameter unitary flow generated by the Hamiltonian; a measurement
updates the state by projection; and composite systems are joined by the
tensor product. Everything else---the quantization of energy, selection
rules, the structure of atomic spectra and of the periodic table---follows by
deduction. The postulates are few, but they are not all of the same
character, and a long line of work has sought to show that some of them are
consequences of the others, or of more primitive principles.

That program is old. Birkhoff and von~Neumann~\cite{birkhoff} sought to read
the Hilbert-space structure off the logic of yes/no measurements; Gleason
\cite{gleason} showed that the Born rule is not independent but forced by the
lattice of projections; Wigner~\cite{wigner} established that symmetries must
be represented by unitary or antiunitary maps; and Stone~\cite{stone} proved
that every strongly continuous one-parameter group of unitaries possesses a
unique self-adjoint generator. More recent reconstructions derive the entire
formalism from operational or informational axioms~\cite{hardy,cdp}. In nearly
all of this the kinematic and probabilistic postulates are the ones placed
under scrutiny; the dynamical postulate---the Schr\"odinger equation---is
usually left standing as an independent assumption. Even when it is motivated
rather than assumed, the motivation runs through Wigner and Stone from the
prior input that evolution is \emph{unitary}: unitarity is supplied by hand,
and the generator follows. The Schr\"odinger equation is, in short, the
postulate of quantum dynamics that is least often derived, and when derived it
rests on unitarity assumed at the outset.

This paper asks what happens if the dynamical postulate is replaced by a
single physical principle that does \emph{not} presuppose unitarity: that
physical time evolution conserves the possible outcomes of measurement. Since
the outcomes of an observable are the points of its spectrum, the principle is
that evolution be \emph{isospectral}. We show that this one demand, imposed on
the Hermitian observables of a Hilbert space, forces the dynamics into Lax
form, $\dot A=[M,A]$; forces the flow to be unitary; and---through Stone's
theorem---produces a unique self-adjoint Hamiltonian as its generator. The
Heisenberg equation, the two Schr\"odinger equations, the conservation laws,
and the good quantum numbers then follow as theorems. The Schr\"odinger
equation is thereby demoted from axiom to consequence, and the dynamical
postulate is replaced by the more elementary statement that measurement
outcomes are conserved in time. The organizing structure throughout is the
Lax pair, the object through which integrable systems express the
conservation of a spectrum.

A Lax pair is a mathematical device that renders a nonlinear evolution exactly
solvable~\citep{lax, prasolov, goriely}: a pair of operators $(\Lop,\Mop)$ for which the dynamics
is equivalent to $\dot\Lop=\comm{\Mop}{\Lop}$, an evolution that moves $\Lop$
only by conjugation and so freezes its spectrum. In the inverse scattering
transform~\cite{ggkm,akns,drazin,faddeev} the conserved spectral data of
$\Lop$ are exactly the quantities that linearize the problem. The historically
decisive instance is the Korteweg--de~Vries equation (KdV)~\cite{ggkm}:  $u_t=6uu_x-u_{xxx}$ , whose Lax
operator is the time-independent Schr\"odinger operator $-\partial_x^2+u$: as
the field $u$ evolves, the bound states (solitons) and the scattering spectrum
of its Schr\"odinger problem are the constants of motion. This is the one classical setting in which the Lax operator and
the quantum spectral problem coincide outright; it is the hinge on which the
remainder of the paper turns.

Such coincidence is usually read in one direction---an integrable field
theory borrows the Schr\"odinger operator as a computational device. We ask
whether the arrow can be reversed. Suppose one is handed nothing but a state
space with the structure of a complex Hilbert space, on which observables are
Hermitian operators whose eigenvalues are the outcomes of measurement, and one
imposes the single structural demand that these observables evolve
isospectrally, preserving their measurement spectra. Does quantum mechanics
follow? In particular, can the Hamiltonian, the Schr\"odinger equation, and
the notion of a good quantum number be \emph{derived} rather than presupposed?
The body of the paper shows that they can.

\section{Lax pairs}
\label{sec:theory}

\subsection{Definition}

Let $\Lop$ and $\Mop$ be linear operators depending on a real parameter
$t$. The pair $(\Lop,\Mop)$ is a \emph{Lax-pair}~\citep{lax, prasolov, goriely} for the evolution of
$\Lop$ if
\begin{equation}
\frac{\dd \Lop}{\dd t} \;=\; \comm{\Mop}{\Lop}
\;=\; \Mop\Lop - \Lop\Mop .
\label{eq:lax}
\end{equation}
The content of~\eqref{eq:lax} is most transparent through the linear
\emph{eigenvalue problem} carried by $\Lop$,
\begin{equation}
\Lop\,\psi = \lambda\,\psi ,
\label{eq:eig}
\end{equation}
together with a linear law fixing the time dependence of its
eigenfunctions,
\begin{equation}
\partial_t \psi = \Mop\,\psi .
\label{eq:auxtime}
\end{equation}
Equations~\eqref{eq:eig}--\eqref{eq:auxtime} form the \emph{auxiliary linear
problem}; their compatibility is exactly the Lax equation~(\ref{eq:lax}), as we now show.

\subsection{Isospectrality}

Differentiate the eigenvalue equation~\eqref{eq:eig} in time, allowing both
$\psi$ and $\lambda$ to depend on $t$, and substitute
$\partial_t\psi=\Mop\psi$:
\begin{align}
\dot\Lop\,\psi + \Lop\Mop\psi
   &= \dot\lambda\,\psi + \lambda\,\Mop\psi
   = \dot\lambda\,\psi + \Mop\Lop\psi ,
\end{align}
where the last step used $\lambda\Mop\psi=\Mop(\lambda\psi)=\Mop\Lop\psi$.
Rearranging,
\begin{equation}
\big(\dot\Lop - \comm{\Mop}{\Lop}\big)\psi = \dot\lambda\,\psi .
\label{eq:isoproof}
\end{equation}
Hence, \emph{if} the Lax equation~\eqref{eq:lax} holds, then
$\dot\lambda=0$: the eigenvalues of $\Lop$ are constants of motion. This is
the defining property---the flow~\eqref{eq:lax} is \emph{isospectral}---and
it has been obtained directly from the eigenvalue problem, with no need to
construct any evolution operator. The same conclusion holds for every
spectral invariant: using the cyclicity of the trace~\citep{lax, prasolov, goriely},
\begin{equation}
\frac{\dd}{\dd t}\,\Tr \Lop^{k}
   = k\,\Tr\!\big(\Lop^{k-1}\comm{\Mop}{\Lop}\big) = 0,
\qquad k=1,2,\dots
\label{eq:invariants}
\end{equation}
so all power sums $\Tr\Lop^{k}=\sum_n\lambda_n^{k}$ are conserved.

\section{Quantum mechanics from spectrum preservation}
\label{sec:construction}

We now suppose given only a state space, its observables, and the bare
existence of a spectrum-preserving time evolution, and derive quantum
mechanics. The Lax equation, the Hamiltonian, and the Schr\"odinger equation
all appear as constructed objects.

\subsection{The primitives}

\paragraph{Axiom A1 (state space).} The state space is a complex Hilbert
space $\Hilb$.

\paragraph{Axiom A2 (observables and measurement).} Observables are
Hermitian operators on $\Hilb$. The possible outcomes of measuring an
observable $\Aop$ are the points of its spectrum
$\spec(\Aop)\subset\mathbb{R}$, and a time-evolved observable is again an
observable.

\paragraph{Axiom A3 (dynamics).} Physical time evolution of the observables
exists and preserves the measurement structure. It is a strongly continuous
one-parameter group of linear maps $\Phi_t$ ($\Phi_0=\mathrm{id}$,
$\Phi_{t+s}=\Phi_t\Phi_s$), under which every observable
$\Aop\mapsto\Aop(t)=\Phi_t(\Aop)$ retains its full spectral data,
\begin{equation}
\spec\Aop(t)=\spec\Aop(0)\qquad\text{(eigenvalues with multiplicities).}
\label{eq:isoaxiom}
\end{equation}
No form is assumed for the evolution beyond continuity, linearity, and the
preservation of spectra.
No Lax equation, Hamiltonian, energy, or Schr\"odinger equation appears in
these axioms. We extract them all, beginning with the Lax equation itself.

\subsection{From isospectral evolution to Lax dynamics}

The Lax equation is not postulated; it follows from Axiom~A3. By Axiom~A2
each evolved observable $\Aop(t)=\Phi_t(\Aop)$ is Hermitian, and by
Axiom~A3 it carries the spectral data of $\Aop(0)$; two Hermitian operators
with the same eigenvalues and multiplicities are unitarily equivalent
(spectral theorem). Moreover $\Phi_t$ is a single linear spectrum-preserving
map of \emph{all} observables, and a linear, spectrum-preserving bijection of
the self-adjoint operators is necessarily a conjugation~\cite{marcusmoyls,jafarian}, so one and the same
operator implements the equivalence for every observable. Hence, in finite
dimensions and locally in $t$,
\begin{equation}
\Aop(t) = G(t)\,\Aop(0)\,G(t)^{-1},\qquad G(0)=\Id,
\label{eq:conj}
\end{equation}
with $G(t)$ common to all observables. Differentiating~\eqref{eq:conj},
\begin{equation}
\dot\Aop(t) = \dot G G^{-1}\,\Aop(t) - \Aop(t)\,\dot G G^{-1},
\end{equation}
that is,
\begin{equation}
\dot\Aop(t) = \comm{\Mop(t)}{\Aop(t)},
\qquad \Mop(t)=\dot G(t)\,G(t)^{-1}.
\label{eq:laxderived}
\end{equation}
The Lax equation is therefore a \emph{theorem}: it is the differential form
of a continuous, spectrum-preserving evolution, and $\Mop$ is the generator
of that flow. The possible measurement values are transported in time
without being changed; Lax dynamics is the preservation of the measurement
structure itself.

The same Hermiticity fixes the nature of $G$ and $\Mop$. Because two
Hermitian operators with equal spectra are related by a \emph{unitary}, the
conjugation~\eqref{eq:conj} may be taken with $G(t)=U(t)$ unitary
(equivalently, imposing $\Aop(t)^\dagger=\Aop(t)$ in~\eqref{eq:conj} forces
$G^\dagger G\propto\Id$, whose residual scalar commutes with every
observable, drops out of~\eqref{eq:laxderived}, and reappears later as the
additive constant in the energy). Differentiating $U^\dagger U=\Id$,
\begin{equation}
\Mop^\dagger = -\,\Mop \qquad(\text{$\Mop$ anti-Hermitian}),
\end{equation}
so $\{U(t)\}_{t\in\mathbb{R}}$ is a strongly continuous one-parameter group
of unitary operators: isospectral evolution of Hermitian observables is
\emph{unitary} evolution. In infinite dimensions the explicit
$U(t)=\ee^{\Mop t}$ and the termwise differentiation are replaced by their
rigorous counterpart---a strongly continuous one-parameter unitary group, to
which Stone's theorem applies directly.

\subsection{Stone's theorem: the Hamiltonian as output}

A strongly continuous one-parameter unitary group has, by Stone's theorem~\cite{stone},
a unique self-adjoint generator. Applied to $\{U(t)\}$ it yields a unique
self-adjoint operator $H$ with
\begin{equation}
U(t) = \ee^{\,\ii H t/\hbar},
\qquad\text{equivalently}\qquad
H \;:=\; -\,\ii\hbar\,\Mop\,.
\label{eq:hamiltonian}
\end{equation}
Self-adjointness is immediate: $H^\dagger=\ii\hbar\,\Mop^\dagger=
-\ii\hbar\,\Mop=H$. The constant $\hbar$ converts the generator (dimension
$1/\text{time}$) into an energy, and the residual scalar freedom noted above
reappears as the additive constant in $H$---the conventional origin of
energy, equivalent to a global phase of $U$.

This is the crux. We did not posit a Hamiltonian; we posited only that a
continuous evolution preserves the measurement spectrum (Axiom~A3) and that
observables remain observables (Axiom~A2). From these the Lax equation, the
unitarity of the flow, and---through Stone's theorem---a self-adjoint
generator were forced into existence. That generator is the Hamiltonian;
being self-adjoint, it is itself an observable, so energy joins the
measurable quantities as a consequence rather than an assumption.

\subsection{The Heisenberg and Schr\"odinger equations}

Substituting~\eqref{eq:hamiltonian} into the Lax
equation~\eqref{eq:laxderived} gives, for every
observable,
\begin{equation}
\frac{\dd \Aop}{\dd t} = \frac{\ii}{\hbar}\comm{H}{\Aop},
\label{eq:heisenberg}
\end{equation}
the Heisenberg equation of motion---now a theorem. Taking the distinguished
Lax operator to be $H$ itself, the eigenvalue problem~\eqref{eq:eig} reads
\begin{equation}
H\psi = E\psi,
\label{eq:tise}
\end{equation}
the time-independent Schr\"odinger equation, with the spectral parameter
identified as the energy, $\lambda=E$, conserved by isospectrality
($\dot E=0$). States, dual to observables, evolve by $G(t)^{-1}=\ee^{-\Mop t}$,
so $\partial_t\psi=-\Mop\psi=-\tfrac{\ii}{\hbar}H\psi$, i.e.
\begin{equation}
\ii\hbar\,\partial_t\psi = H\psi,
\label{eq:tdse}
\end{equation}
the time-dependent Schr\"odinger equation. The discrete and continuous
parts of $\spec(H)$ are the bound states (the quantization condition) and
the scattering sector, the roles played by solitons and radiation for the
KdV equation.

\subsection{Good quantum numbers as stationary Lax operators}

An observable is conserved iff it is a stationary point of the flow:
\begin{equation}
\frac{\dd \Aop}{\dd t}=0
\;\Longleftrightarrow\;
\comm{\Mop}{\Aop}=0
\;\Longleftrightarrow\;
\comm{H}{\Aop}=0 .
\label{eq:stationary}
\end{equation}
The conserved observables are precisely the stationary Lax operators; their
eigenvalues are constant labels, fixed by the state and simultaneously
diagonalizable with $H$. These are the \emph{good quantum numbers}. A
maximal commuting family $\{C_1,\dots,C_r\}$ with $\comm{H}{C_a}=0$
generates a tower of mutually commuting isospectral flows,
$\partial_{s_a}\Aop=\tfrac{\ii}{\hbar}\comm{C_a}{\Aop}$, with
$[\partial_{s_a},\partial_{s_b}]=0$; its joint spectrum
$\{(E,c_1,\dots,c_r)\}$ is the complete set of good quantum numbers
labelling the stationary states.

\subsection{Constants of motion}
\label{sec:constants}

The framework yields conserved quantities of two complementary kinds:
\emph{stationary Lax operators}---observables that commute with the
generator and therefore do not move at all---and \emph{spectral
invariants}, the scalars $\Tr\Lop^k$ that stay constant even when the Lax
operator itself flows. The two are unified by a single criterion: the
power traces $\Tr X^k$ are conserved precisely when $X$ evolves by
conjugation, i.e.\ obeys a Lax equation of von Neumann--Heisenberg form
$\dot X=\pm\tfrac{\ii}{\hbar}\comm{H}{X}$ (with explicit time dependence
permitted). Stationary operators are the special solutions with
$\comm{H}{X}=0$, for which $X^k$ itself is constant; the flowing solutions
conserve only the traces. We give examples of each.

\paragraph{Symmetry invariants (stationary Lax operators).} Whenever $H$ is
invariant under a symmetry with self-adjoint generators $\{T_a\}$, each
generator commutes with $H$ and is a stationary Lax operator,
$\dot T_a=\tfrac{\ii}{\hbar}\comm{H}{T_a}=0$; the Casimir invariants built
from them are simultaneously diagonalizable with $H$, and their conserved
eigenvalues are the good quantum numbers labelling the irreducible
multiplets. For a rotationally invariant (central) Hamiltonian the
components $L_i$ generate $\mathfrak{so}(3)$ with quadratic Casimir
$L^2=\sum_i L_i^2$; the stationary Lax operators $L^2$ and $L_z$ carry the
conserved eigenvalues $\hbar^2\ell(\ell+1)$ and $\hbar m$, so that the
complete set $\{H,L^2,L_z\}$ labels the stationary states $\ket{n\ell m}$.
Translational invariance makes each momentum component $p_i$ stationary,
$\dot{\mathbf p}=0$, with conserved eigenvalue $\hbar\mathbf k$; reflection
symmetry makes parity $\Pi$ ($\Pi^2=\Id$) a stationary Lax operator
with eigenvalue $\pm1$; and a number-conserving many-body Hamiltonian makes
the particle number $N$ stationary, with conserved integer eigenvalue. The
framework registers \emph{hidden} symmetries on the same footing: for the
Coulomb Hamiltonian $H=\mathbf p^2/2m-k/r$ the Laplace--Runge--Lenz vector
\begin{equation}
\mathbf A=\frac{1}{2m}\big(\mathbf p\times\mathbf L-\mathbf L\times\mathbf p\big)
          -k\,\hat{\mathbf r}
\label{eq:rungelenz}
\end{equation}
satisfies $\comm{H}{\mathbf A}=0$ although it generates no geometric
symmetry; together with $\mathbf L$ it closes into $\mathfrak{so}(4)$, and
its constancy is precisely the accidental $\ell$-degeneracy of the hydrogen
spectrum~\cite{pauli}. Every observable commuting with the generator is thus a constant
of the Lax flow, dynamical and geometric symmetries alike, and its
eigenvalues are good quantum numbers.

\paragraph{The Hamiltonian (spectral invariants).} The
invariants~\eqref{eq:invariants} of $\Lop=H$ are the energy moments
$I_k=\Tr H^k=\sum_n E_n^k$, whose generating functional is the partition
function
\begin{equation}
Z(\beta) = \Tr\,\ee^{-\beta H}
        = \sum_{k\ge 0}\frac{(-\beta)^k}{k!}\,I_k ,
\end{equation}
itself a Lax invariant. Equilibrium statistical mechanics is the bookkeeping
of the conserved spectral data of the quantum Lax operator.

\paragraph{The density operator (spectral invariants).} A state is described by a
density operator $\rho$, which evolves by the von Neumann equation. Written
with $H=-\ii\hbar\Mop$,
\begin{equation}
\frac{\dd\rho}{\dd t} = -\frac{\ii}{\hbar}\comm{H}{\rho}
                       = \comm{-\Mop}{\,\rho\,},
\label{eq:vonneumann}
\end{equation}
which is itself a Lax equation~\eqref{eq:lax}, with $\rho$ as the Lax
operator and the state-side generator $\tilde\Mop=-\Mop$. Unlike the
symmetry invariants above, $\rho$ is in general \emph{not} stationary---it
flows---yet the flow is isospectral: the eigenvalues of $\rho$ (the
populations $p_n$) are conserved, and so is every $\Tr\rho^k$. In particular the purity and the
von Neumann entropy,
\begin{equation}
\Tr\rho^2 = \sum_n p_n^2,
\qquad
S = -\Tr(\rho\ln\rho) = -\sum_n p_n\ln p_n,
\end{equation}
are constants of motion. The reversibility of closed quantum evolution---the
constancy of purity and entropy---is the isospectrality of the density
operator's Lax flow, the same property that conserves the energy spectrum,
applied now to the state. Because every function of an isospectral operator
shares its conserved spectrum, the entire family of R\'enyi entropies
\begin{equation}
S_\alpha = \frac{1}{1-\alpha}\,\ln\Tr\rho^{\alpha}
\qquad (\alpha>0),
\end{equation}
is conserved, with the von Neumann entropy recovered as $\alpha\to1$. The
restriction to \emph{closed} evolution is essential: a reduced density
operator $\rho_A=\Tr_B\rho$ evolves non-unitarily, is \emph{not}
isospectral, and its spectral invariants---hence the entanglement
entropy---do change in time. Isospectrality is a property of the global
flow alone.

\paragraph{Dynamical invariants for driven systems (spectral invariants).}
The von Neumann form $\dot X=-\tfrac{\ii}{\hbar}\comm{H}{X}$ continues to
define an isospectral flow when the generator is time dependent,
$H=H(t)$. A Hermitian operator $I(t)$ obeying
\begin{equation}
\frac{\partial I}{\partial t} = -\frac{\ii}{\hbar}\comm{H(t)}{I}
\label{eq:lewis}
\end{equation}
is a \emph{dynamical invariant} in the sense of Lewis and Riesenfeld~\cite{lewis}: its
eigenvalues are constant in time although both $I(t)$ and $H(t)$ are not,
so every $\Tr I^k$ is conserved and its eigenstates (carrying
Lewis--Riesenfeld phases) solve the time-dependent Schr\"odinger equation.
The density operator is the canonical solution
of~\eqref{eq:lewis}; the nontrivial ones make driven problems tractable.
For the time-dependent oscillator
$H(t)=p^2/2m+\tfrac12 m\omega^2(t)x^2$ the invariant is
\begin{equation}
I(t)=\frac12\!\left[\Big(\frac{x}{b}\Big)^2+\big(b\,p-m\dot b\,x\big)^2\right],
\label{eq:erminv}
\end{equation}
where $b(t)$ solves the Ermakov--Pinney equation~\cite{pinney}
$\ddot b+\omega^2(t)\,b=1/(m^2 b^3)$; the conserved eigenvalues of $I$ are
$(n+\tfrac12)\hbar$, the oscillator ladder frozen even as $\omega(t)$
varies. A projector onto an evolving state,
$P(t)=U(t)\ket{\psi_0}\bra{\psi_0}U(t)^\dagger$, is the degenerate special
case with frozen spectrum $\{1,0,\dots\}$. Thus, beyond the symmetry
charges and the static spectrum of $H$, the operators whose power traces are
constants of motion are exactly the dynamical invariants---the density
operator and its functions for autonomous evolution, and the
Lewis--Riesenfeld invariants for driven evolution.

\section{Conclusion}

We have recast quantum mechanics as a Lax-pair theory and, in doing so,
reduced its axiomatic basis. Three postulates suffice: a complex Hilbert
space of states (Axiom~A1), Hermitian observables whose spectra are the
outcomes of measurement (Axiom~A2), and the demand that a continuous time
evolution preserve those spectra (Axiom~A3). From these the Lax form of the
dynamics is a theorem---a spectrum-preserving evolution of Hermitian
observables is necessarily a unitary conjugation $\dot\Aop=\comm{\Mop}{\Aop}$
with $\Mop$ anti-Hermitian---and Stone's theorem then delivers a unique
self-adjoint operator, the Hamiltonian, as the generator of the resulting
unitary group.
The Heisenberg equation, the time-dependent and time-independent
Schr\"odinger equations, the conservation laws, and the good quantum numbers
all follow as theorems. In particular, the Schr\"odinger equation and the
unitary time evolution it defines---postulated outright in the conventional
formulation---are here not assumed but derived. The dynamical postulate of
quantum mechanics has been removed from the foundations.

The reach of the construction should be stated precisely. What it
reconstructs is the dynamical and spectral architecture of the theory,
together with the Hamiltonian itself as an output rather than an input. What
it does not replace is the kinematic and interpretive core: the
Hilbert-space structure (Axiom~A1), the Hermiticity of observables
(Axiom~A2), and the Born rule connecting spectral projections to
probabilities remain the measurement postulates that the present formulation
shares with the standard one. The claim is therefore sharp---it is the
\emph{dynamics}, the Schr\"odinger postulate, that is shown to be superfluous
as an axiom, not the kinematics or the probability rule.

Two further points delimit the result. First, the Lax flow obtained here is
universally isospectral, a property of unitary conjugation alone; it should
not be conflated with Liouville integrability, the stronger requirement that
a sufficiently large family of commuting observables exist to label the
states, which only special systems possess. Second, the analytic
ingredients---Stone's and Wigner's theorems, and the fact that the time
evolution is a one-parameter automorphism group of the observable
algebra---are classical; the novelty lies in the axiomatic reorganization,
in which the Lax structure is taken as primitive and the Schr\"odinger
equation descends from it.

The conserved quantities of the theory appear, in this language, as the
spectral data of Lax operators: the symmetry invariants, among them the
angular momentum $L^2$ and the hidden Runge--Lenz vector, as the stationary
Lax operators; the energy moments and the partition function as spectral
invariants of $H$; and the purity, R\'enyi, and von Neumann entropies---and,
for driven systems, the Lewis--Riesenfeld invariants---as spectral invariants
of the density operator and its generalizations. In its dynamical and
spectral content, quantum mechanics is Lax-pair theory, formulated with one
postulate fewer than before.


\end{document}